# Improvement of Solid-Fluid Interaction Scheme in Lattice Boltzmann Immiscible Pseudopotential Models

Yizhong Chen (陈怡忠) [1,2], Zhibin Wang (王智彬) [1,2*], Shi Tao (陶实)[3*], Zhi Yang(杨智) [1,2], Ying Chen (陈颖)[1,2]

[1]*School of Material and Energy, Guangdong University of Technology, Guangzhou 510006, China*

[2]*Guangdong Provincial Key Laboratory of Functional Soft Matter, Guangzhou 510006, China*

[3]*School of Chemical Engineering and Energy Technology, Dongguan University of Technology, Dongguan 523808, China*

*Corresponding author:

[1,2,]Tel: + 86 020-39322570; E-mail: wangzhibin@gdut.edu.cn (Zhibin Wang).

[3,]Tel: + 86 0769-22862039; E-mail: taoshi@dgut.edu.cn (Shi Tao).

## Abstract

The pseudopotential lattice Boltzmann method (LBM) is a prominent approach in computational fluid dynamics, valued for its physical intuitiveness and computational tractability. However, existing solid-fluid interaction schemes for modeling wettability in multi-component immiscible fluid systems face persistent challenges, notably spurious currents and unphysical mass-transfer layers. This study proposes an improved solid-fluid interaction scheme that preserves implementation simplicity while effectively mitigating these numerical artifacts. Furthermore, leveraging the superior isotropy of eighth-order discrete schemes, we develop a novel boundary treatment methodology for second-layer lattice data reconstruction at complex interfaces. To validate the universality of the proposed improved scheme, four benchmark scenarios are simulated: static contact angle measurement on cylindrical surface, droplet dynamics through confined geometries, immiscible displacement processes, and co-current flow in microchannels. Results demonstrate that the improved scheme accurately captures diverse complex immiscible multiphase flows.

## NOMENCLATURE

| | |
|---|---|
| $\boldsymbol{f}$ | density distribution function |
| $\boldsymbol{f}^{eq}$ | equilibrium density distribution function |
| $\boldsymbol{F}$ | external forces term |
| $\boldsymbol{u}$ | velocity, lattice units (lu/ts) |
| $\boldsymbol{u}^{eq}$ | common velocity, lattice units (lu/ts) |
| $w$ | weight coefficient |
| $\boldsymbol{e}$ | discrete velocity |
| $\boldsymbol{F}_{k_in}$ | intermolecular forces between two fluids, lattice units (mu·lu/ts$^2$) |
| $G_{12}$ | strength coefficient of interaction between two fluids |
| $\boldsymbol{F}_{k_ads}$ | adhesive forces between fluid and solid, lattice units (mu·lu/ts$^2$) |
| $G_w$ | strength coefficient of interaction between solid and fluid |
| $\boldsymbol{F}_v$ | volume force, lattice units (mu·lu/ts$^2$) |
| $x$ | *X-axis* of Cartesian coordinate system |
| $y$ | *Y-axis* of Cartesian coordinate system |
| $z$ | *Z-axis* of Cartesian coordinate system |
| $M_v$ | viscosity ratio |
| $M_\rho$ | density ratio |

## Greek symbols

| | |
|---|---|
| $\tau$ | relaxation time |
| $\upsilon$ | kinematic viscosity, lattice units (lu$^2$/ts) |
| $\rho$ | density, lattice units (mu/lu$^3$) |
| $\rho^e$ | effective density |
| $\psi$ | pseudopotential function |
| $\delta_t$ | time step, lattice units (ts) |
| $\Omega$ | collision operator |
| $\boldsymbol{I}$ | unit second-order tensor |

## Superscripts/subscripts

| | |
|---|---|
| $k$ | index of different fluid components |
| $i$ | lattice directions |
| $m$ | main component |
| $d$ | dissolved component |
| $a$ | analytical solution |

Some symbols with special meanings are described separately in the text.

## I. Introduction

As a mesoscopic numerical approach rooted in kinetic theory, the lattice Boltzmann method (LBM) has established itself as an efficient computational framework for simulating complex fluid flow and heat transfer[1–5]. Notably in multiphase flow simulations, the kinetic nature of LBM provides distinct advantages in interfacial handing through automatic phase interface maintenance, effectively circumventing the intricate interface tracking procedures required by conventional numerical methods like Volume of Fluid and Level-Set techniques[1,6,7]. Among the diverse LBM multiphase formulations developed over the past decades[8–11], the pseudopotential model distinguishes itself through its elegant physical interpretation and computational simplicity. This innovative approach conceptualizes interphase interactions via a density-dependent pseudopotential mechanism, where spontaneous phase separation emerges naturally from carefully designed short-range attractive forces[12–14]. The model's intrinsic ability to capture interfacial phenomena without complex surface tension calculations has propelled its widespread adoption in simulating multiphase systems ranging from droplet dynamics to porous media flows.

Wetting phenomena pervade diverse scientific and engineering domains, ranging from microfluidic devices to geological processes[15–17]. At the heart of wettability characterization lies the contact angle ($\theta$), a fundamental metric quantifying the interfacial tension equilibrium at the triple junction where liquid-solid and liquid-fluid interfaces converge. A surface is deemed hydrophilic when $\theta < 90°$ and hydrophobic when $\theta > 90°$, with these thresholds dictating fluid spreading behavior through Young's equation. In pseudopotential LBM implementations, precise wettability control necessitates carefully designed solid-fluid interaction schemes. The pioneering work by Martys and Chen[18] established a seminal adhesion force framework through parameterized solid-fluid interactions (governed by adhesion parameter $G$), enabling systematic contact angle modulation. This elegant yet effective methodology has become ubiquitous in droplet dynamics studies across hydrophilic/hydrophobic surfaces[19–22], though its practical limitations became apparent through pronounced spurious currents near boundaries that compromise numerical stability[23]. Subsequent refinements addressed these computational artifacts: Raiskinmaki et al.[24,25] introduced pseudopotential-function-modified interactions that mitigated spurious currents in single-component gas-liquid systems. However, their contact angle model is based on gas-liquid phase systems, where there is a nonlinear coupling relationship between the pseudo potential function and fluid density, for example the exponential function form ($\varphi(\rho) = e^{(-1/\rho)}$) or equation of state (such as P-R, C-S equations). This coupling mechanism can effectively distinguish the characteristic differences of different phases. But for immiscible multiphase systems, since there is no considering phase transition process, it is only necessary to make the pseudopotential function equal to the fluid density. Therefore, it is not feasible to reduce unphysical effects such as spurious velocity by modifying the pseudopotential function for immiscible two-phase systems. Kang et al.[26] then devised a dual-parameter scheme combining wall density $\rho_w$ and adhesion strength *G*. This scheme partially suppressed spurious currents while extending applicability to immiscible systems, albeit with constrained angular range ($45° < \theta < 135°$). However, due to the need to adjust two parameters simultaneously and continuously try to obtain the desired contact angle, this method lacks convenience in implementation compared to the original scheme. Parallel developments by Colosqui et al.[27] proposed a mesoscopic derived interaction potential with repulsive-attractive components, achieving enhanced interfacial dynamics fidelity through rigorous kinetic theory foundations. Nevertheless, its computational complexity and single-component limitation present formidable challenges for immiscible multiphase extensions.

Critical analysis of existing methodologies reveals persistent challenges in pseudopotential-based immiscible multiphase simulations: (1) predominant solid-fluid interaction schemes remain largely confined to single-component systems, (2) current implementations for immiscible fluids suffer from restricted wettability ranges ($\theta \approx 33°$-$120°$) when using original solid-fluid scheme, and (3) inherent trade-offs persist between numerical stability and implementation complexity[26,28]. This methodological gap becomes particularly critical when simulating practical multiphase systems requiring precise yet robust wettability control across broad contact angle spectra. This study seeks to overcome through our proposed methodology.

Our proposed advancement addresses these limitations through a novel solid-fluid interaction paradigm integrating mechanical stability principles with localized adhesion modulation. Building upon Li's interfacial stabilization framework[28,29], we introduce a dynamic adhesion parameter $G(x)$ that adapts to local effective densities of immiscible phases, fundamentally re-engineering traditional static schemes. This formulation achieves three critical improvements demonstrated in subsequent validation: (1) 33% reduction in average spurious currents, (2) extended contact angle range ($\theta \approx 15°$-$165°$), and (3) maintained computational simplicity through single-parameter calibration.

Implementation validation employs a hierarchical verification protocol: (1) Static wettability characterization on complex geometries, (2) Dynamic multiphase scenarios, and (3) Quantitative benchmarking against analytical solutions and prior numerical studies. The scheme's elimination of unphysical mass-transfer artifacts confirms enhanced mechanical stability near boundaries. Theoretical significance emerges from the scheme's unique capacity to decouple wettability control from phase separation dynamics. Practical implications extend to high-stability simulations in porous media flows and microscale fluidic networks where conventional methods exhibit prohibitive numerical diffusion. The proposed eighth-order isotropy treatment with adaptive boundary reconstruction further enhances numerical robustness for complex geometries, achieving <1.84% velocity profile deviation in benchmark tests. Finally, it should be emphasized in this paper that all units are based on lattice units (dimensionless), making the research results more universal. Additionally, since the N-S equation can be converted into a dimensionless equation and its simulation parameters also can be dimensionless numbers, thus, in most flows, if the key dimensionless criterion numbers are kept the same, similar flow fields can be obtained. In general, in the LBM simulation, using similarity criteria to ensure that the main criteria number matches the problem does not require unit conversion to restore the values of all variables to their actual physical units[29].

## II. Pseudopotential Multiphase Model of LBM

### A. Theoretical Framework

LBM operates at the mesoscopic scale, resolving fluid dynamics through discrete distribution functions $f_i^k(x,t)$ that evolve via collision and streaming processes on a structured lattice. For immiscible two-phase systems, the evolution equations for each component $k \in \{1, 2\}$. This kinetic-theory-based approach inherently captures interfacial dynamics in multiphase systems without explicit interface tracking, making it particularly suited for immiscible fluid simulations[30].

For immiscible two-phase flows, the pseudopotential model employs dual sets of distribution functions $(f_i^1, f_i^2)$ to represent distinct fluid components. Each component adheres to the discrete Boltzmann equation:

$$\boldsymbol{f}_i^k(x+\boldsymbol{e}_i\Delta t, t+\Delta t)-\boldsymbol{f}_i^k(x,t)=\Omega_i^k+(1-\frac{1}{2\tau_k})\Delta t\cdot\boldsymbol{F}_i^k(x,t)\,. \tag{1}$$

The collision operator $\Omega_i^k$ adopts the single-relaxation-time (SRT) approximation:

$$\Omega_i^k=\frac{\boldsymbol{f}_i^{eq,k}(x,t)-\boldsymbol{f}_i^k(x,t)}{\tau_k}\,. \tag{2}$$

Once the distribution function for each lattice point is obtained, the macroscopic density and velocity at that location can be calculated:

$$\rho_k=\sum_i f_i^k, \tag{3}$$

$$\rho_k u_k=\sum_i f_i^k e_i, \tag{4}$$

$$u^{eq}=\frac{\sum_k\rho_k u_k}{\sum_k\rho_k}. \tag{5}$$

In the pseudopotential model, each lattice point has two density values, the main density and the dissolved density. However, the "dissolved" referred to here is an algorithmic phenomenon in the pseudopotential model, not the actual physical dissolution[30]. For example, at a lattice point where fluid 1 is the main component, then the main density at that point is the density of fluid 1. The dissolved density is the density of fluid 2. As shown in Fig. 1(a) and 1(b), the main density ($\rho_k^m$) characterizes the dominant fluid within the lattice point, while the dissolved density ($\rho_k^d$) represents the mathematical presence of another fluid in the corresponding region.

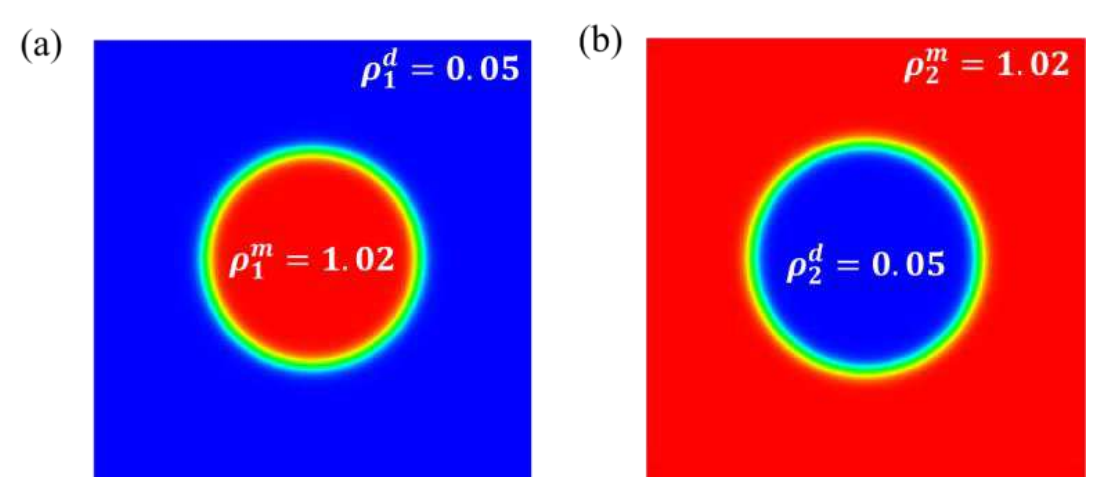

**Fig. 1**. Density contour plots of fluid 1 (a) and fluid 2 (b).

## B. Evolution Equations and Force Coupling

For pseudopotential models, the equilibrium density distribution function is like that of single phase and the external force term adopts the Guo's format[31] in this paper.

The details are as follows:

$$\boldsymbol{f}_i^{eq,k} = w_i \rho_k [1 + \frac{\boldsymbol{e}_i \cdot \boldsymbol{u}^{eq}}{c_s^2} + \frac{(\boldsymbol{e}_i \cdot \boldsymbol{u}^{eq})^2}{2c_s^4} - \frac{\boldsymbol{u}^{eq} \cdot \boldsymbol{u}^{eq}}{2c_s^2}]. \quad (6)$$

External forces term:

$$\boldsymbol{F}_i^k(x,t) = w_i [\frac{\boldsymbol{e}_i \cdot \boldsymbol{F}_k}{c_s^2} + \frac{(\boldsymbol{u}^{eq}\boldsymbol{F}_k + \boldsymbol{F}_k \boldsymbol{u}^{eq}) : (\boldsymbol{e}_i\boldsymbol{e}_i - c_s^2 \boldsymbol{I})}{2c_s^4}]. \quad (7)$$

It is embedded in Eq. (1) through Eq. (7). The details of intermolecular forces are shown below, and adhesive forces are introduced in the next section along with the improved scheme proposed in this paper:

$$\boldsymbol{F}_{k_in} = -\psi_k(x) G_{12} \sum_i \psi_k(x + \boldsymbol{e}_i)\boldsymbol{e}_i \,. \quad (8)$$

The magnitude of $G_{12}$ controls the surface tension between the immiscible fluids. Therefore, in the actual simulation, it is important to select a large enough value to ensure the incompatibility of the two-phase fluid.

## C. Solid-Fluid Interaction Schemes

### 1. Original Scheme

To simulate the interaction between fluid and solid walls, an adhesion force must be introduced to accurately describe their contact behavior. Martys and Chen[18] proposed the following solid-fluid interaction scheme:

$$\boldsymbol{F}_{k_ads} = -G_{w,k} \rho_k(x) \sum_i w_i s(x + \boldsymbol{e}_i \Delta t)\boldsymbol{e}_i \,. \quad (9)$$

$G_{w,k}$ represents the adhesive parameter of the different fluid components, $s(x + e_i \Delta t)$ is switch function that equals 0 for fluid lattice points and 1 for solid lattice points. By adjusting $G_{w,k}$, different contact angles can be achieved. In this original solid-fluid interaction scheme, it uses a constant adhesive parameter $G_{w,k}$ across all computational grid points near the solid surface to calculate adhesive force. This means the adhesion strength between fluid and solid is uniformly applied everywhere, regardless of local physical variations. However, according to Li's[32] theory of mechanical stability, that is maintaining a sharp phase interface, such uniform treatment could create unrealistic effects. When the same adhesion parameter is forced at every location, it may distort the fluid's equilibrium state near the wall[23].

**2. Improved Scheme**

To address the generation of spurious currents, we propose an improved solid-fluid interaction scheme that uses a local adhesive parameter instead of the constant interaction parameter. Therefore, we construct effective density ( $\rho^e$ ) and calculate the corresponding local interaction parameter.

$$\rho^e = \frac{\rho_1(x,t) - \rho_2(x,t)}{\rho_1(x,t) + \rho_2(x,t)}, -1 \le \rho^e \le 1. \tag{10}$$

After obtaining the macroscopic density of each fluid component ( $\rho_1$ and $\rho_2$ ), the effective density at the corresponding lattice point can be obtained by Eq. (10). It is used to identify the dominant component at this lattice point. And then, $G_{w,1}$ can be calculated by using Eq. (11) based on the effective density at each lattice point.

$$G_{w,1} = \begin{cases} G_{w,1}, & \rho^e > \delta \\ g_1\ (\rho^e), & \delta \ge \rho^e > 0 \\ g_2\ (\rho^e), & 0 \ge \rho^e > -\delta \\ G_{w,2}, & \rho^e < -\delta \end{cases}. \tag{11}$$

Due to the immiscible pseudopotential model treating the phase interface as a diffusion interface method, there is a certain width of the phase interface for different surface tension. The function of $\delta$ is to achieve a smooth transition in the value of the interaction strength parameter at different interface widths, avoiding stability degradation caused by sudden numerical changes. In this paper, the interaction strength parameter between the two phases is taken as 0.17, and the width of the phase interface is about 5 lattice grids. When $\delta$=0.8, the interval width that can be processed is also around 5 grid widths. Therefore, this paper sets the value of $\delta$ to 0.8. $g_1(\rho^e) = s_1 + s_2\rho^e + s_3(\rho^e)^2$ and $g_2(\rho^e) = t_1 + t_2\rho^e + t_3(\rho^e)^2$ are interpolation function for $\rho^e$. $s_1 = t_1 = 2\frac{G_{w1} \cdot G_{w2}}{G_{w1} + G_{w2}}$, $s_2 = 2\frac{G_{w1} - s_1}{\delta}$, $s_3 = -\frac{s_2}{2\delta}$, $t_2 = 2\frac{t_1 - G_{w2}}{\delta}$, $t_3 = \frac{t_2}{2\delta}$.

As observed, Eq. (11) is a piecewise function. This formulation is motivated by the continuously changing density at the phase interface in pseudopotential model; the piecewise function effectively captures capture this dynamic behavior[33,34]. According to Huang et al.[30]'s research, the parameter $G_{w,2}$ can be set to $G_{12}$, therefore, it just needs to specify the upper bound of the piecewise function by assigning a value to $G_{w,1}$. As previously discussed, distinct wetting characteristics are determined by the relative magnitudes of $G_{w,1}$ and $G_{w,2}$. To more intuitively represent the differences between them, we introduce a strength coefficient: $\lambda$ ($\lambda = G_{w1} / G_{w2}$). A surface is considered hydrophilic when $\lambda$ is less than 1, and hydrophobic otherwise. After this change, the corresponding expressions are as Eq.(12) and others parameters change as $g_1^{'}(\rho^e) = s_1^{'} + s_2^{'}\rho^e + s_3^{'}(\rho^e)^2$, $g_2^{'}(\rho^e) = t_1^{'} + t_2^{'}\rho^e + t_3^{'}(\rho^e)^2$, $s_1^{'} = t_1^{'} = \frac{2\lambda}{1+\lambda}$, $s_2^{'} = 2\frac{(\lambda^2 - \lambda)}{\delta \cdot (1+\lambda)}$, $s_3^{'} = \frac{\lambda - \lambda^2}{\delta^2 \cdot (1+\lambda)}$, $t_2^{'} = 2\frac{\lambda - 1}{\delta \cdot (1+\lambda)}$, $t_3^{'} = \frac{\lambda - 1}{\delta^2 \cdot (1+\lambda)}$. For code implementation details, please refer to **Appendix A**.

$$G_{w,1} = G_{w,2} \cdot \begin{cases} \lambda\rho^e > \delta \\ g_1^{'}(\rho^e), \delta \geq \rho^e > 0 \\ g_2^{'}(\rho^e), 0 \geq \rho^e > -\delta \\ \rho^e < -\delta \end{cases}. \quad (12)$$

## D. High-Order Isotropy Implementation

As outlined above, we have discussed the implementation steps for optimizing the original solid-fluid scheme. Additionally, since the order of isotropy can also significantly affect the magnitude of the spurious currents[3,35,36]. To avoid the influence caused by isotropic difference, the isotropy of $8^{th}$ is used[21,37,38] for both original and improved scheme, as shown in the Fig. 2(a). However, it should be noted that special attention needs to be paid to the handling of boundaries (velocity or solid boundaries) when using isotropy larger than $4^{th}$ order. For example, when we use the $8^{th}$ order isotropy, the interaction range spans two lattices, this causes the fluid node at the boundary to appear "empty lattice", that is, it is out of the calculation area, as shown in Fig. 2(b1). In response to this problem, some scholar proposed to further lay a ghost node on the outside of the boundary, so as to meet the information acquisition at the node[2,39,40], as shown in Fig. 2(b2). This method is relatively simple to implement; however, it limits the application of the format when the boundary is a non-straight interface. To better compensate for this shortcoming, this paper proposes an approach of copying fluid node information to deal with the problem of missing information on the second layer node, this method can effectively deal with non-straight surfaces. As seen in Fig. 2(a), the position of second layer lattice points correlates with that of the first layer. Therefore, the information at the second-layer points can be assigned from corresponding the first-layer points. For the curve boundaries, such as point B in Fig. 2(b3), the information of nodes (eg., 14, 22, 18) is unknown. Using the proposed method, this unknown information is assigned from node 6, thus reducing gradient calculation error of macroscopic quantities on the solid surface and enhancing simulation accuracy. Table I describes the first-layer nodes corresponding to each node on the second layer. Whether the fluid node is in velocity boundary or near irregular solid surface, this method adapts well to these situations. In the following discussion, this paper will further prove the feasibility of this method.

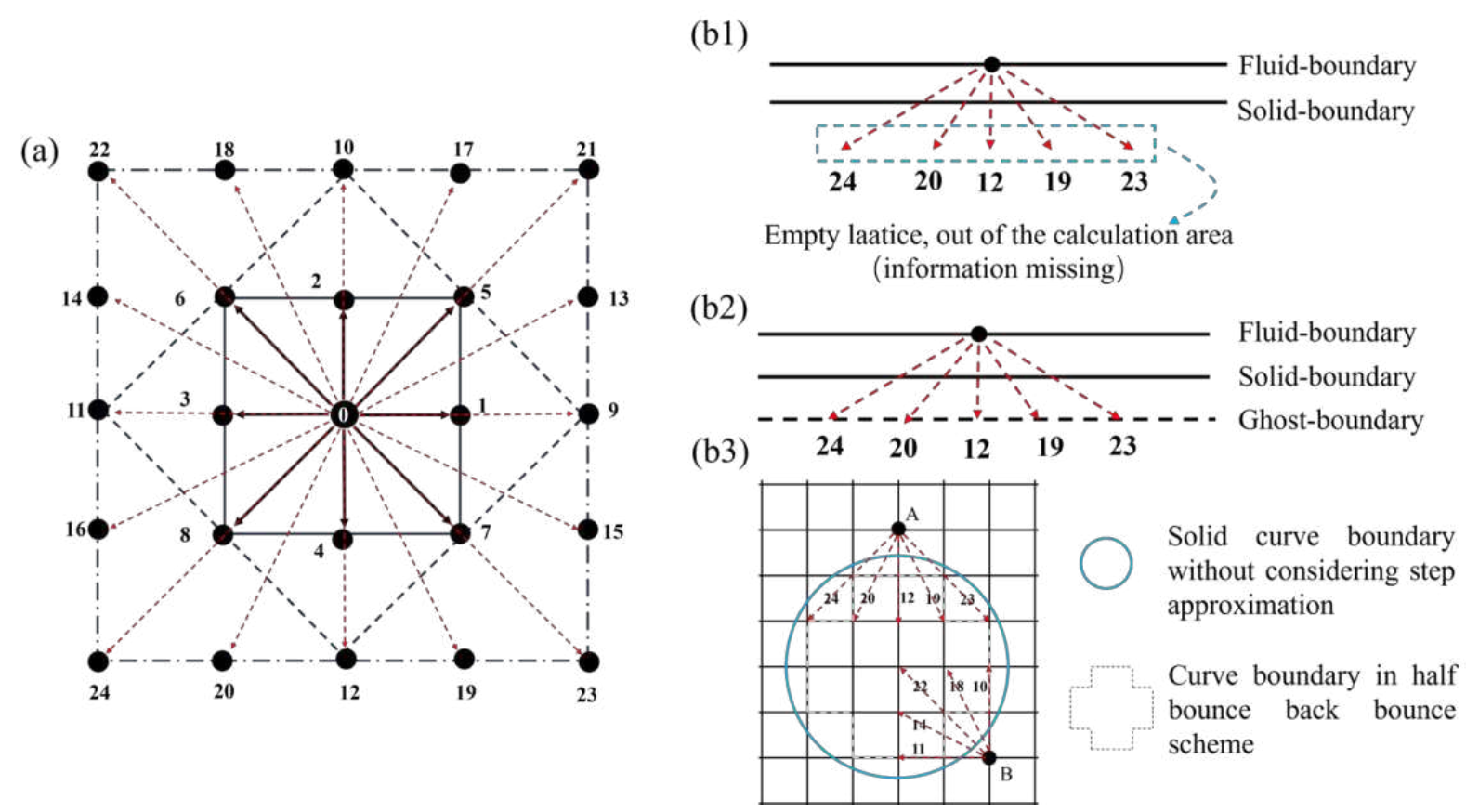

**Fig. 2**. Multi-range interaction scheme for 8th-order isotropy. (a) 24 lattice positions with two-layer interactions. (b1) Lattices layers beyond the calculation area. (b2) Ghost-node layer reconstructed to supply missing information. (b3) Schematic diagram of the method for curve boundary.

**Table I**. Node correspondence.

| Second-layer nodes ($n$) | 9-16 | 17-20 | 21-24 |
|---|---|---|---|
| First-layer nodes | $n$-8 | $n$-12 | $n$-16 |

## III. Results and discussion

### A. Co-current flow for immiscible fluid

Co-current flow of immiscible fluids is a fundamental benchmark for validating numerical algorithms in multiphase systems, as interfacial errors can propagate and distort macroscopic flow behavior[41]. To rigorously evaluate the accuracy of the improved solid-fluid interaction scheme, we simulate both Poiseuille flow (pressure-driven) and Couette flow (shear-driven) scenarios and compare its analytical solution with our numerical solution. The schematic diagrams of these validation models are shown in Fig.3(a). The computational domain is a 50×100 grid. The top and bottom boundaries employ non-slip conditions (details of the boundary schemes are provided in **Appendix B**), while the lateral boundaries are periodic. Fluid 1 ($\rho_1^m = 1.02$, $\rho_1^d = 0.05$, $\upsilon_1 = 0.17$) is located at $0 \le |y| \le a$ ($a$=25), and fluid 2 ($\rho_2^m = 1.02$, $\rho_2^d = 0.05$, $\upsilon_2 = 0.17$) is located at $a \le |y| \le L$ ($L$=50). When the absolute difference in velocity between 10,000 times steps is less than $10^{-5}$, we consider the simulation results to converge.

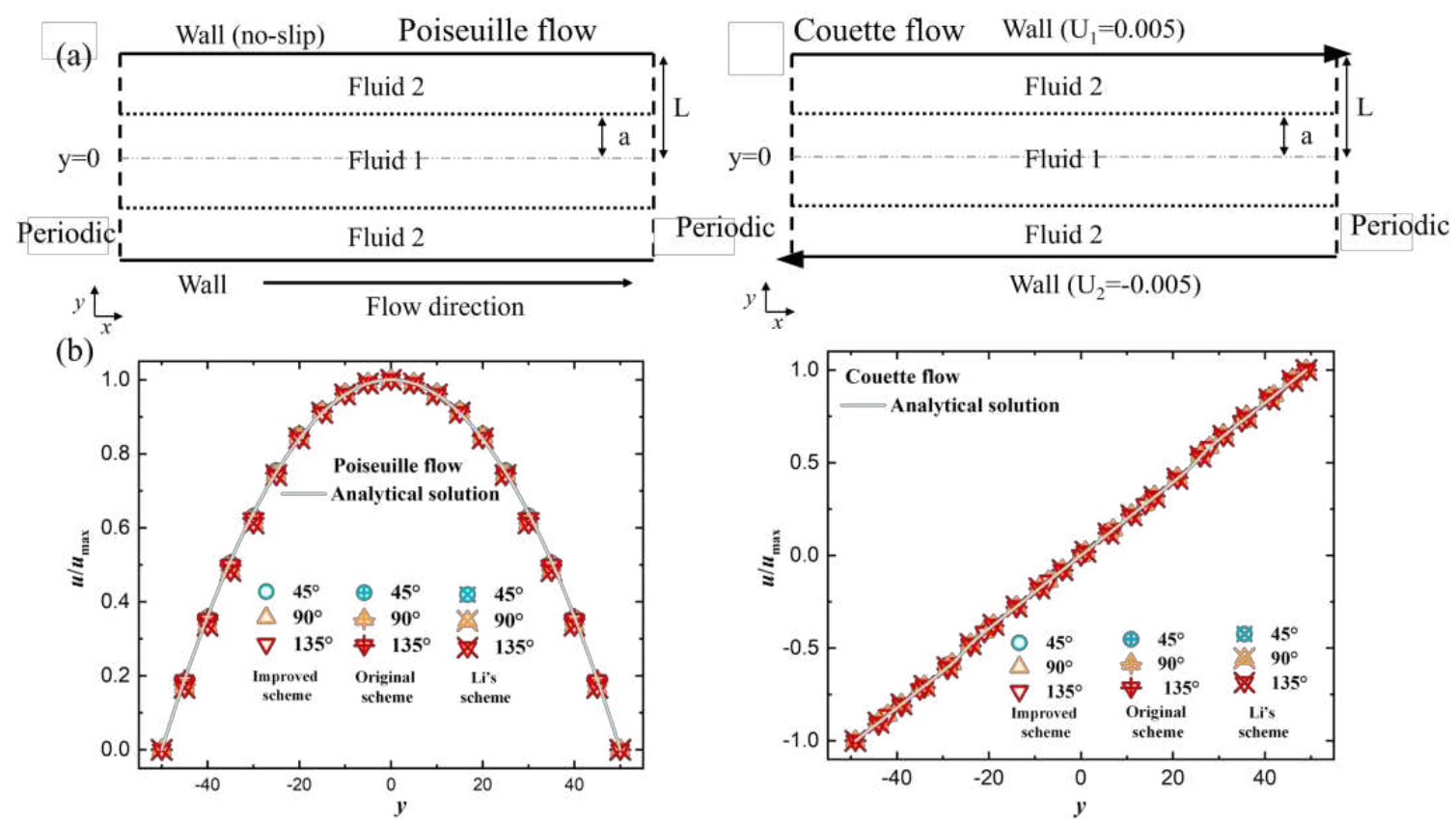

**Fig. 3.** Model and validation of Poiseuille and Couette flow. (a) Schematic diagrams. (b) Velocity profiles compared with analytical solution at different contact angles.

In the case of Poiseuille flow, the horizontal volume force $F_{1_v} = F_{2_v} = 10^{-6}$ is used to drive the fluid movement to represent the pressure gradient in the horizontal direction of the Poiseuille flow. For Couette flow, the top and bottom walls are set as sliding boundaries with constant horizontal velocity ($U_1$=-$U_2$=0.005). The analytical solutions for both Poiseuille and Couette flow are Eq. (13.a) and Eq. (13.b), respectively.

$$u_x^a(y)_{\text{poiseuille}} = \begin{cases} \dfrac{F_b}{2\upsilon_2\rho_2}(L^2 - y^2), & a \le |y| \le L \\ \dfrac{F_b}{2\upsilon_1\rho_1}(a^2 - y^2) + \dfrac{F_b}{2\upsilon_2\rho_2}(L^2 - a^2), & 0 \le |y| \le a \end{cases} . \quad (13.a)$$

$$u_x^a(y)_{\text{couette}} = \begin{cases} 2y, & a \le |y| \le L \\ y+1, & 0 \le |y| \le a \end{cases} . \quad (13.b)$$

Fig. 3(b) compares the normalized velocity profiles from simulations using the improved scheme with analytical solutions across three contact angles: $\theta$ = 45°, 90°, and 135°. The simulated profiles align closely with analytical solutions for both flow types, with relative errors ($\varepsilon$) consistently below 1.84%. The improved scheme accurately captures flow dynamics regardless of wettability, demonstrating its robustness in diverse interfacial configurations. Convergence is achieved within 10,000 iterations (velocity difference < $10^{-5}$), underscoring the scheme's computational efficiency. The well agreement between analytical and numerical solutions strictly proves the accuracy of the improved solid-fluid scheme proposed in this paper. Furthermore, we rigorously compare the three contact angle schemes. Quantitative evaluation reveals minimal discrepancies in their interfacial dynamics simulation results. This finding indicates that the proposed improved scheme preserves the fundamental hydrodynamic characteristics at solid-fluid interfaces while maintaining functional equivalence to exiting schemes. Consequently, it enables reliable characterization of surface wettability phenomena. Therefore, the subsequent discussion focuses on exploring the improved scheme's performance across various application scenarios.

### B. Contact Angles on Cylindrical Surface

To verify the ability of the proposed improved solid-fluid interaction scheme to simulate different contact angles, we perform simulations of a droplet resting on a cylindrical surface at varying contact angles. In these simulations, two immiscible fluids with equal viscosity ($\upsilon_1 = \upsilon_2$=0.17), main density ($\rho_1^m = \rho_2^m$=1.02) and dissolved density ($\rho_1^d = \rho_2^d$=0.05) were used. The strength of the interaction between the two phases was set to $G_{12}$ = 0.17, which is sufficient to ensure phase separation and immiscibility. The computational domain was a 300×350 grid, with a solid cylinder of radius $R$=70 positioned at (150, 130). A circular droplet (fluid 1, radius $r$=50) was placed at (150, 230), and the remaining area was filled with fluid 2. The domain was set with periodic boundary conditions around the perimeter, and the cylindrical surface was assigned a no-slip condition. The schematic diagram of the initial state is shown as Fig. 4. Additionally, Appendix C provides relevant conversion information between the actual sizes and lattice units.

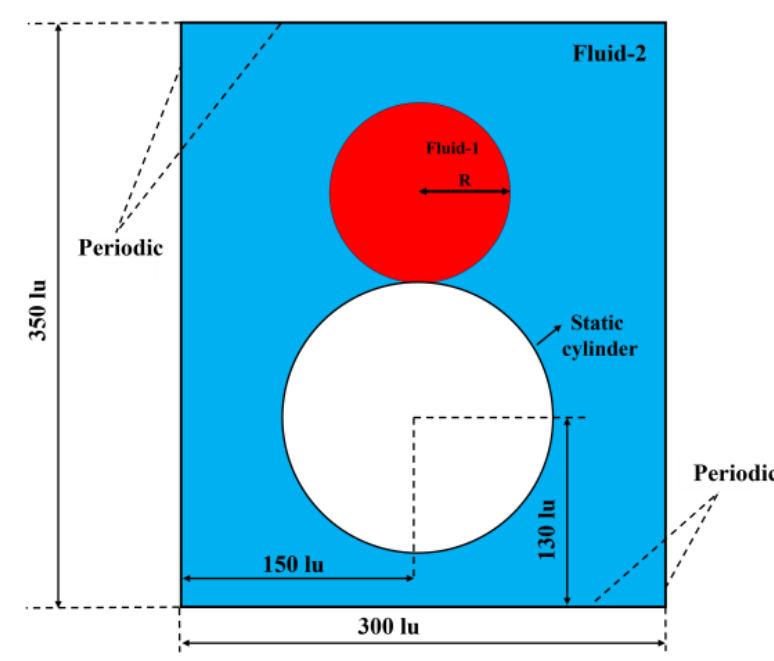

**Fig. 4**. Schematic of the initial droplet distribution on a cylindrical surface.

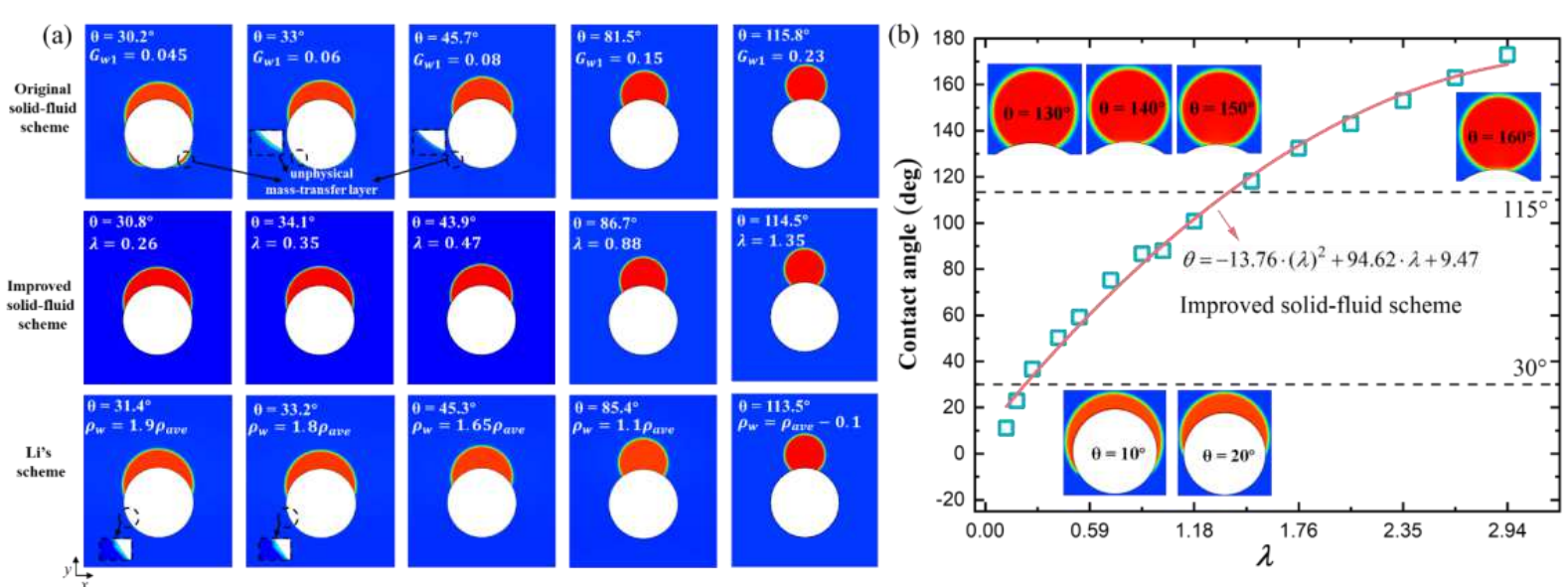

**Fig. 5**. Verification of static wetting behavior. (a) Static contact angles obtained by different schemes. (b) Contact angles under different strength coefficient ( $\lambda$ ).

Fig. 5(a) compares the static contact angles simulated by the original and improved solid-fluid schemes. Both approaches successfully reproduce varied contact angles on a cylindrical surface through parameter adjustment. However, as evidenced in Fig. 5(a), the original scheme exhibits unphysical mass-transfer layer near the solid boundary when the contact angle is small, such as $\theta = 30.2°$. These interfacial anomalies intensify with further reduction of contact angles. In contrast, the improved solid-fluid scheme eliminates such an unphysical mass-transfer layer. This improvement stems from a critical modification: while the original scheme employs a constant interaction strength parameter, compromising mechanical stability at dissolution density interfaces, the improved approach implements spatially variable interaction parameters localized to solid boundaries. This adaptive parameterization preserves interfacial stability by dynamically adjusting interaction strengths, thereby suppressing the unphysical mass-transfer phenomena observed in the original model. Furthermore, Fig. 5(a) also compared improved scheme with Li's scheme[46], showing close agreement between both simulation results. However, Li's scheme[46] exhibits a thin unphysical mass transfer layer near the solid surface at small contact angles (~30°), whereas the improved scheme eliminates this artifact. This demonstrates that the improved scheme not only suppresses spurious mass-transfer layers but also accurately implements diverse wetting characteristics. To facilitate practical implementation, we also established a predictive correlation between contact angles and solid-fluid interaction strength parameters, $\theta = -13.76 \cdot \lambda^2 + 94.62 \cdot \lambda + 9.47$, as shown in Fig. 5(b).

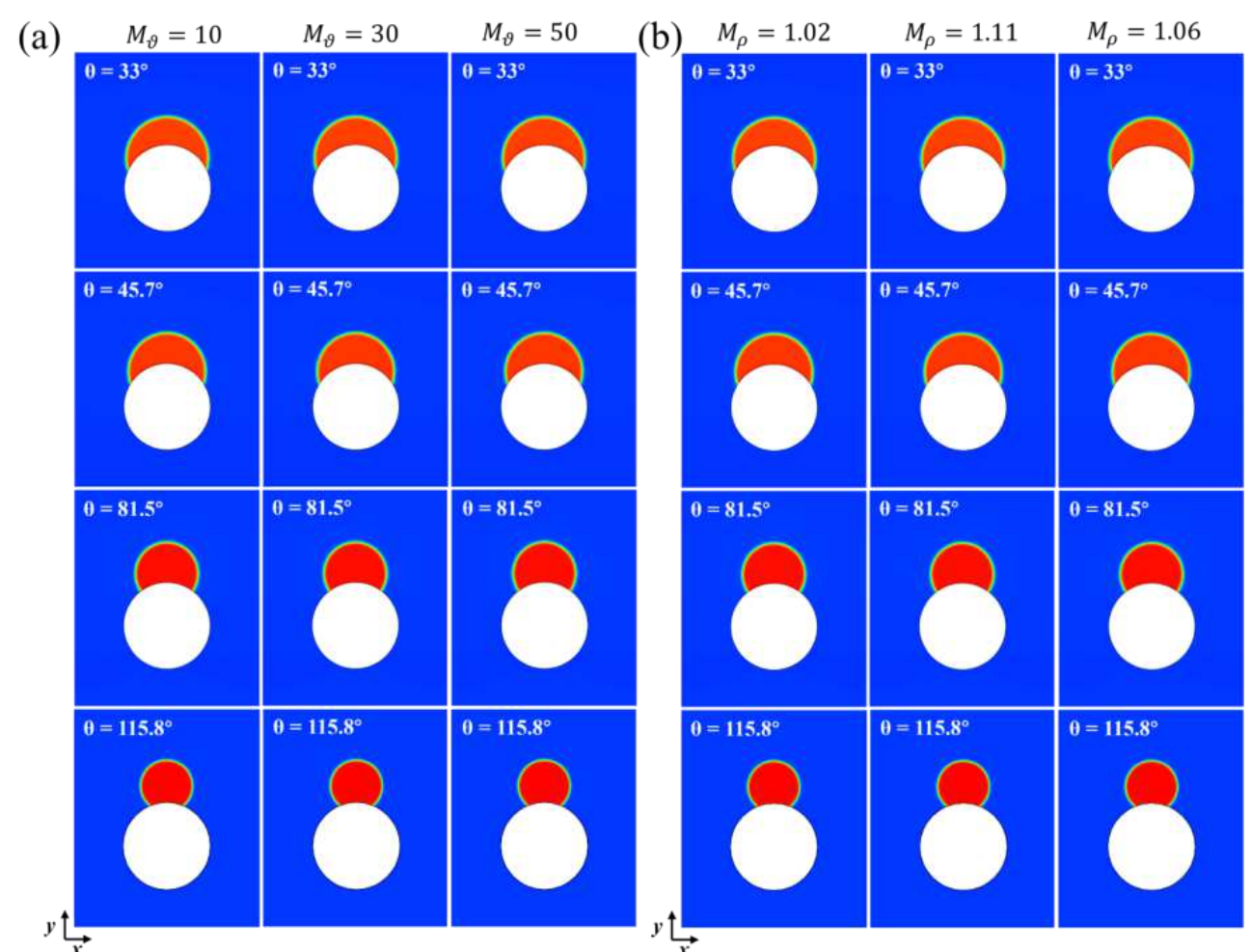

**Fig. 6.** Adaptability of improved scheme. (a) Under different viscosity ratios. (b) Under different density ratios.

To further verify the universality of the improved scheme in the application field, this paper also tested the adaptability of the improved scheme under different viscosity and density ratios (*M* is the ratios of the physical properties of fluid 1 to fluid 2), as shown in Fig. 6. The graph demonstrates consistent simulated contact angles across viscosity and density ratios, with no additional errors attributable to physical property variations. This further demonstrates the universality and stability of the improved scheme proposed in this paper. Given the immiscible pseudopotential model's inherent limitation of optimal stability near unity density ratio (≈1) and its primary applications in microscale systems—where gravity effects are negligible, as in porous media displacement[42] and microfluidic droplet generation[19,43]—we set the two-phase density ratio to 1 is reasonable.

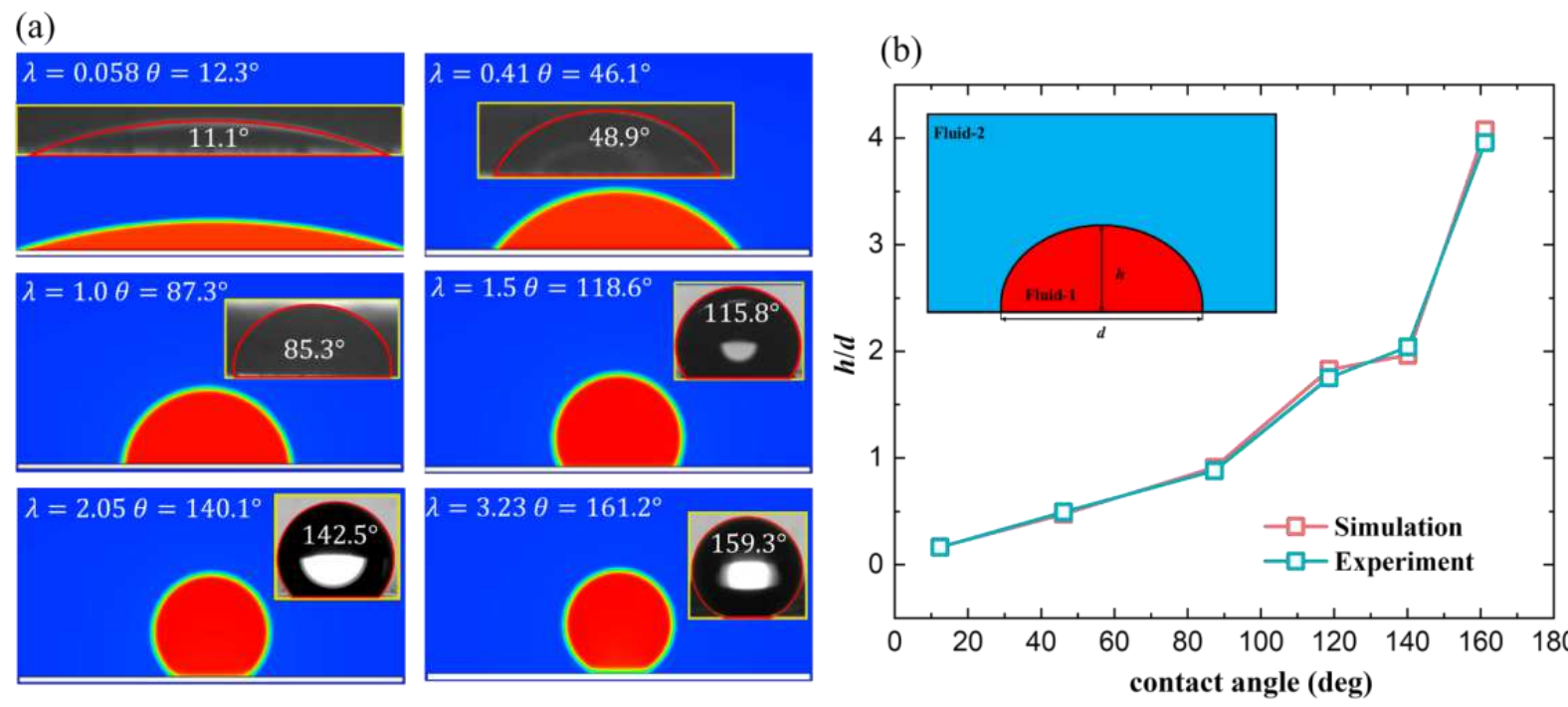

**Fig. 7.** Comparison of simulation and experimental results on flat. (a) Contact angle versus strength coefficient. (b) Droplet height-to-diameter ratio (*h*/*d*) versus contact angle.

Although the preceding validation focused on curved geometries, the improved scheme advanced herein remains equally effective for planar walls. To illustrate this claim, six configurations, each exhibiting distinct droplet wettability on flat substrates, were simulated and compared with the experimental results, as shown in Fig. 7(a) and the experimental details are provided in **Appendix D**. The simulations reproduce the measured equilibrium contact angles with a mean absolute deviation of only 2.1° when compared to experimental results. Additionally, to quantify morphological fidelity further, the height-to-diameter ratio (*h*/*d*) of the static droplet was evaluated (Fig. 7(b)). Owing to the non-negligible gravitational effect in the experiments, the discrepancy increases at larger contact angles. Nevertheless, the overall relative error remains below 5 %, well within acceptable limits. Collectively, the results of Figs. 5–6, together with the experimental comparisons, confirm that the improved scheme successfully eliminates the unphysical mass-transfer layer reported by Sbragaglia et al.[3], even though curved boundaries could not be validated experimentally under the present conditions.

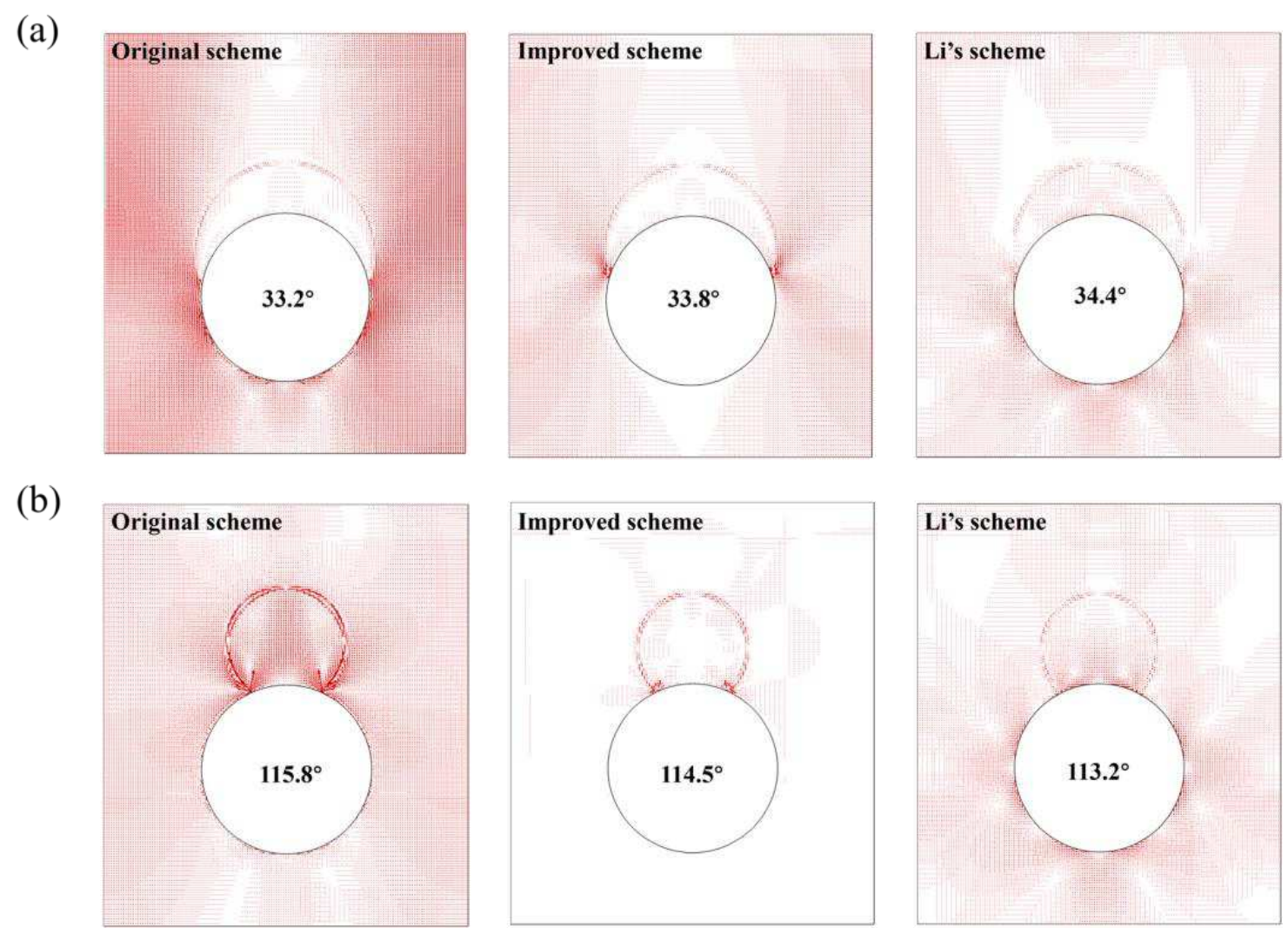

**Fig. 8.** Spurious currents distribution. (a) Hydrophilic surface. (b) Hydrophobic surface.

Spurious currents persist in nearly all numerical simulations of multiphase flow and interface phenomena. Excessive magnitudes can cause severe interfacial distortion. Fig. 8 compares spurious current distributions under hydrophilic ($\theta \approx 30°$) and hydrophobic ($\theta \approx 115°$) conditions for both original and improved solid-fluid interaction schemes. Relative to the original and Li's scheme[46], the improved scheme significantly attenuates spurious currents near solid surfaces by applying localized force corrections in the droplet's non-contact region. This force localization also explains the scheme's efficacy in eliminating unphysical mass transfer layers. However, the localized effects create force imbalances across the contact line, generating localized current maxima. Crucially, these contact-line currents induce negligible disturbance to the global flow field, as evidenced in Fig. 8.

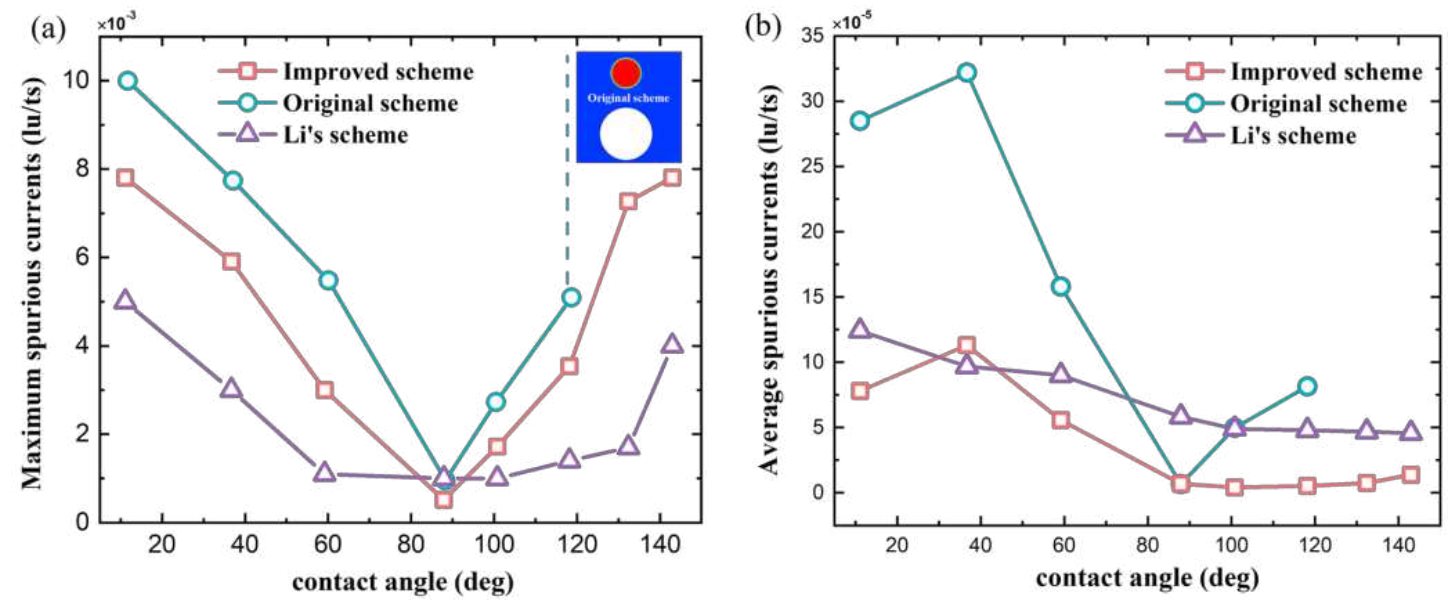

**Fig. 9.** Evaluation of spurious current magnitude for three schemes. (a) Maximum values. (b) Average values.

To quantify numerical performance, Fig. 9 compares the maximum and average spurious currents across the three schemes. The original scheme exhibits significantly higher magnitudes than the improved scheme, suggesting its propensity to generate unphysical mass-transfer layers under low contact angles. Both schemes minimize spurious currents near $\theta \approx 90°$ where the contact line adhesion forces approach equilibrium[44,45]. The improved scheme's localized parameter implementation effectively suppresses spurious currents in hydrophilic regimes. Similarly, its maximum currents remain lower than the original scheme under hydrophobic conditions although the extent of improvement has declined. This decline in improvement may arise from localized adhesion disparities during the convergence of opposing contact lines, which could compromise mechanical stability[23,46]. Crucially, the original scheme fails to sustain droplet at $\theta > 120°$, as illustrated in Fig. 9 (a). This reveals critical limitations in high-contact-angle simulations. Whereas the improved scheme eliminates this inconsistency and doubles the operational range. Comparison with Li's scheme[46] reveals relatively large maximum spurious currents, attributable to force imbalances at the contact line induced by the improved scheme's localized approach. However, the improved scheme achieves a 64% reduction in average spurious currents relative to Li's scheme[46], as shown in Fig. 9(b). This stems from its elimination of unphysical mass-transfer layers, which weakens spurious current distributions system-wide — particularly advantageous when body forces are present.

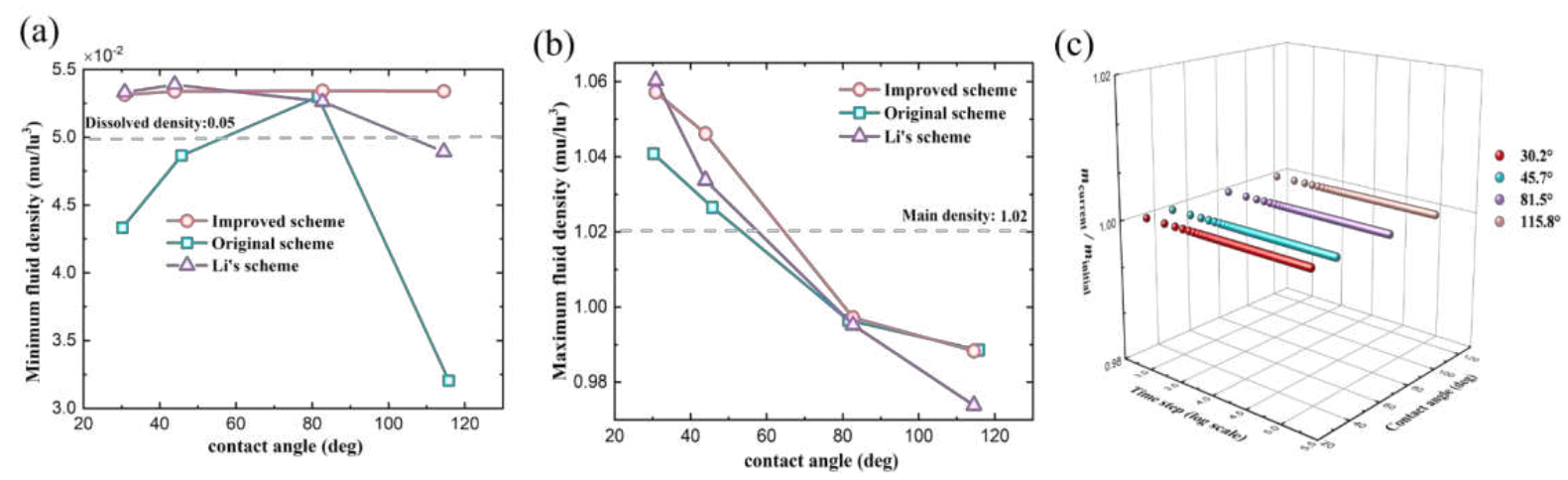

**Fig. 10.** Evaluation of numerical stability and mass conservation. (a) Minimum fluid densities. (b) Maximum fluid densities. (c) Mass conservation performance of the improved scheme.

Numerical disturbances during adhesive force implementation are quantified monitoring deviations of maximum/minimum densities from system-specified reference values[23]. As shown in Fig. 10(a), the improved scheme maintains minimum densities within 0.5%, which is closest to the dissolved density, and there is no significant fluctuation across contact angles. This demonstrates exceptional stability. In stark contrast, the original scheme exhibits significant minimum density fluctuations, aligning with the instability reported by Li et al.[23] and explaining its failure at large contact angles. Comparing the improved scheme with Li's scheme[46], the minimum-density differences between the two schemes are negligible at contact angles below 80°. However, Li's scheme[46] induces significant density fluctuations when angles exceeding 80°. For maximum densities (Fig. 10(b)), the improved scheme yields values slightly higher (approximately 1.02 times) than the original scheme when $\theta < 80°$, but both schemes produce comparable results when $\theta > 80°$. Instead, Li's scheme[46] reduces the density of the solid surface layer when $\theta > 80°$. At contact angles around 120°, this density decreases to approximately 85% of the bulk fluid density. Such a substantial density difference between the solid surface layer and the adjacent fluid leads to a more pronounced impact on the overall fluid density. In summary, we believe that the improved scheme has the best numerical stability across the contact angles. Finally, to ensure the conservation of mass in the improved scheme, the mass ratio ($m_{\text{current}}/m_{\text{initial}}$) was evaluated across different contact angles, as shown in the Fig. 10(c). The results demonstrate that the improved scheme maintains strict mass conservation.

### C. Droplet Dynamics through Confined Geometries

To further validate the accuracy of the improved scheme in complex flow condition, we simulated droplet dynamics through square obstacle in microchannel. The computational domain is 350×150 grid. The square obstacle has a side length of 25 and its center position is (122, 75). At the initial time, the droplet (fluid 1) radius is 25 and its center position at (55, 75), and the rest of area are filled with fluid 2. The top and bottom boundaries of the microchannel are no-slip condition, the inlet is velocity boudnary using Zou-He scheme[5] and outlet boundary is outflow scheme[47]. To avoid the entrance effect, a poiseuille velocity profile $u_{\text{inlet}} = -4U_m y(y-H)/H^2$ ( $U_m$ is maximum value, $H$ is the height of microchannel) is enforced at inlet boundary. The physical properties of fluid 1 and fluid 2 are consistent with **Sec. III A**. During the simulation, the static contact angle of the square-obstacle surface is 30.2°. The schematic diagram of the initial state is shown in Fig. 11(a).

Fig. 11(b) illustrates the results obtained from the original and improved solid-fluid interaction schemes and Li's scheme[46]. The motion attitude of the droplet using original solid-fluid scheme is basically similar to that of the improved scheme and the Li's scheme[46] when $t \leq 6000\delta_t$. But it results in a significant unphysical mass-transfer layer near the surface of the square obstacle compared with other two schemes. Under the influence of the unphysical mass-transfer layer, the droplet on the square obstacle left increases with time, and small droplets also appear on the top and bottom sides of the square obstacle, as illustrated in $t = 10000\delta_t$. In addition, a series of unphysically formed droplets can be found on the top and bottom walls of the microchannel, which is seriously deviated from the real physical results. However, this phenomenon is notably absent in both the improved and Li's scheme[46]. The simulation results of the two schemes are in well agreement. In summary, the improved scheme can accurately capture droplet dynamics through square obstacle and effectively avoid unphysical phenomena.

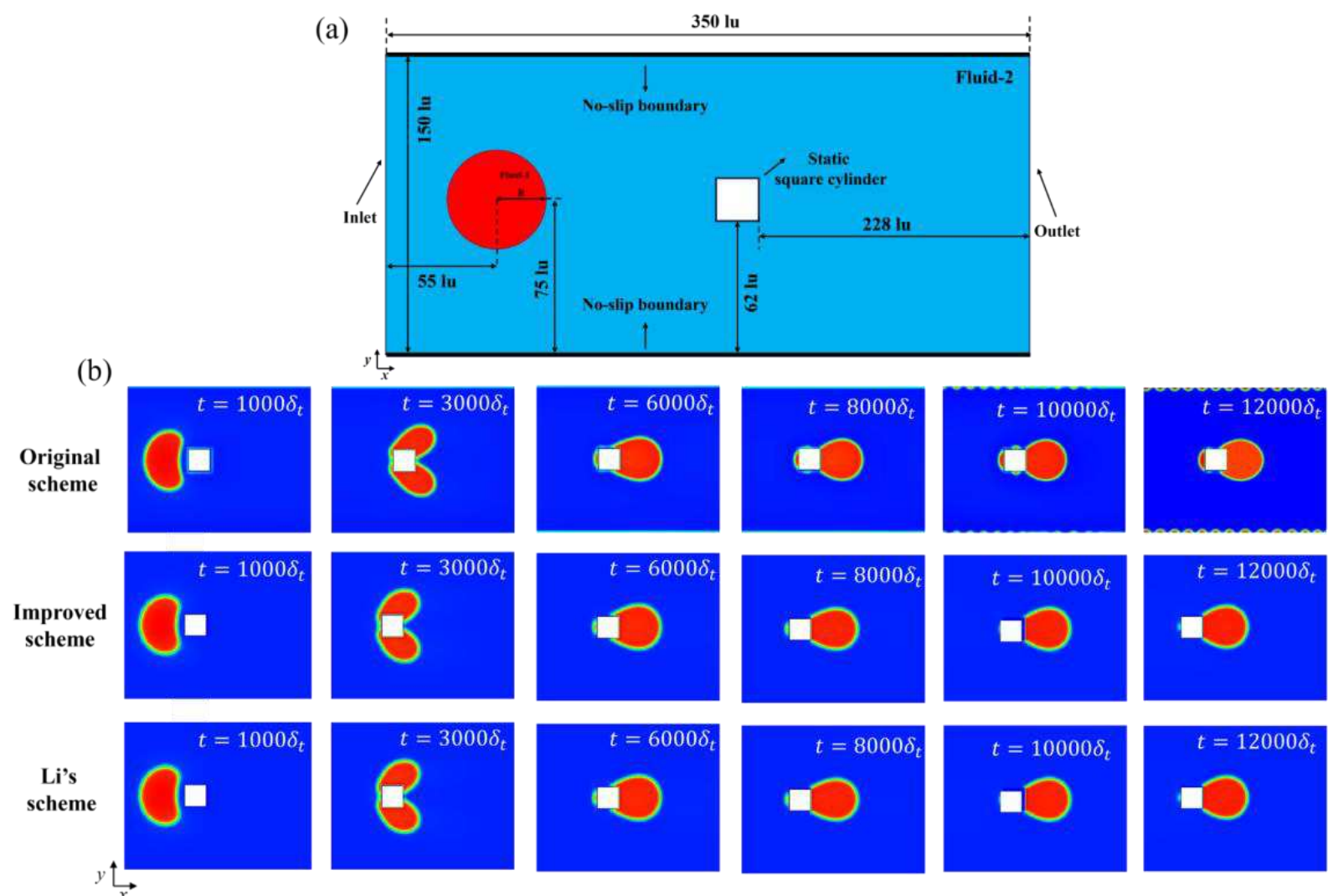

**Fig. 11.** Model and its validation. (a) Schematic diagram of initial droplet in confined geometries. (b) Comparison of schemes for droplet impact on a square surface (θ ≈ 30.2°) at Re = 7.5.

## D. Immiscible Displacement in Microchannel

Immiscible displacement phenomena are critical in understanding multiphase flow dynamics[42,48,49], particularly in applications such as enhanced oil recovery and microfluidic device design. To further validate the universality of the improved solid-fluid interaction scheme, we simulate immiscible displacement in a microchannel and compared with the results of Kang et al.[50]. The computational domain spans 400 × 66 lattice units with boundary conditions and fluid properties consistent with that of Kang et al.[50]. Initially, fluid 1 occupies the first *x*-column, while the remaining region is filled with fluid 2, as shown in Fig.12(a). And Fig. 12(b) shows various quantitative parameters during the process of viscous fingering.

According to Fig. 12(c), the displacement process presented in this paper aligns qualitatively with the observations of Kang et al.[50]. To quantitatively assess the accuracy of the improved scheme, we calculated finger length ($T$), finger width ($W$), and slip distance ($S$) and normalized ($T/L$, $S/L$, and $W/H$) them for comparison with the results of Kang et al.[50], as shown in Table II. The relative errors between our results and the results of Kang et al.[50] are below 5%, confirming the scheme's robustness in capturing interfacial dynamics. In general, from the comparison results with the original solid-fluid interaction schemes, the improved scheme has well accuracy.

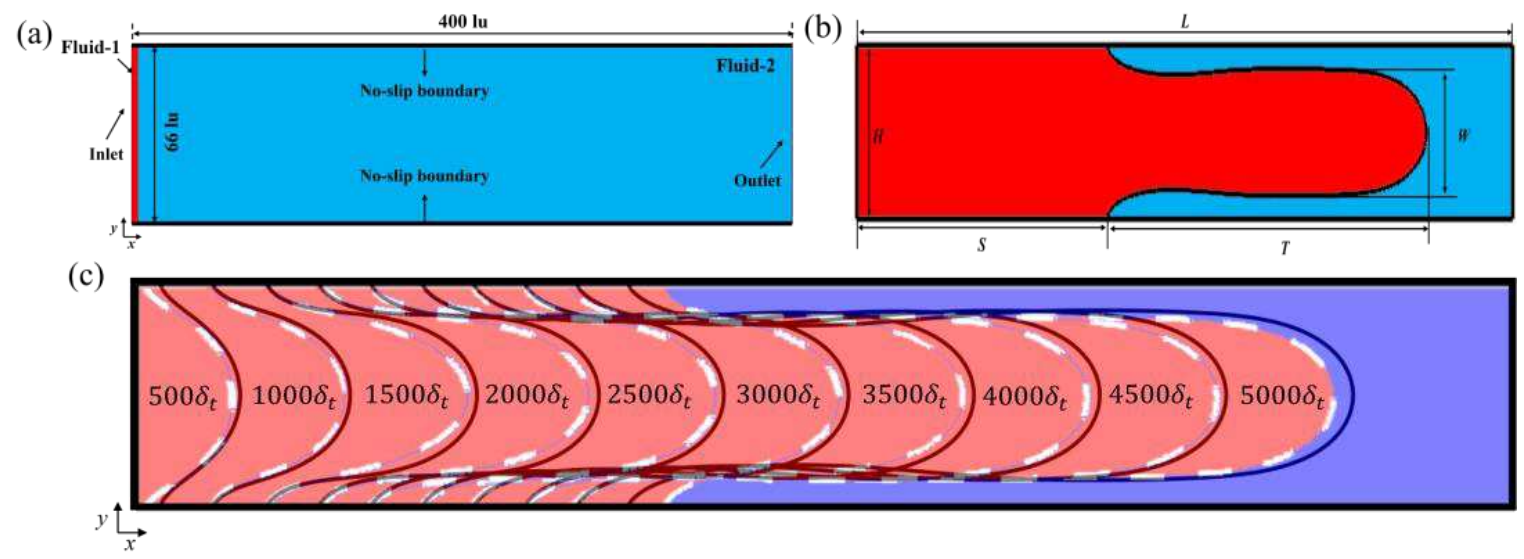

**Fig. 12.** Model and its validation: (a) Schematic diagram of initial state. (b) Phenomenon of viscous fingering. (c) Comparison of interface shapes simulated by improved scheme (white dashed) and Kang et al.[50] (black solid).

**Table II**. Dimensionless finger length (*T*/*L*), slip distance (*S*/*L*), and finger width (*W*/*H*) and the corresponding relative errors.

| | Results of Kang et al.[50] | Improved scheme | Relative error |
|---|---|---|---|
| *T*/*L* | 0.527 | 0.507 | 3.7% |
| *S*/*L* | 0.358 | 0.375 | 4.7% |
| *W*/*H* | 0.743 | 0.775 | 4.3% |

To further test the adaptability of the improved scheme to strong non-uniform flow, this paper added two columns of cylindrical arrays based on this microchannel displacement to simulate the scenario of porous media displacement and compared and verified it with Li's scheme[46], as shown in Fig.13(a) and (b). At different times, the displacement effect exhibited by the improved scheme is almost identical to that of Li's scheme[46]. Additionally, it is worth noting that there is a significant unphysical density gradient on the solid surface in Li's scheme[46], which doesn't occur in the improved scheme. This further demonstrates the superior inhibitory effect of the improved scheme on the unphysical mass transfer layer on solid surfaces.

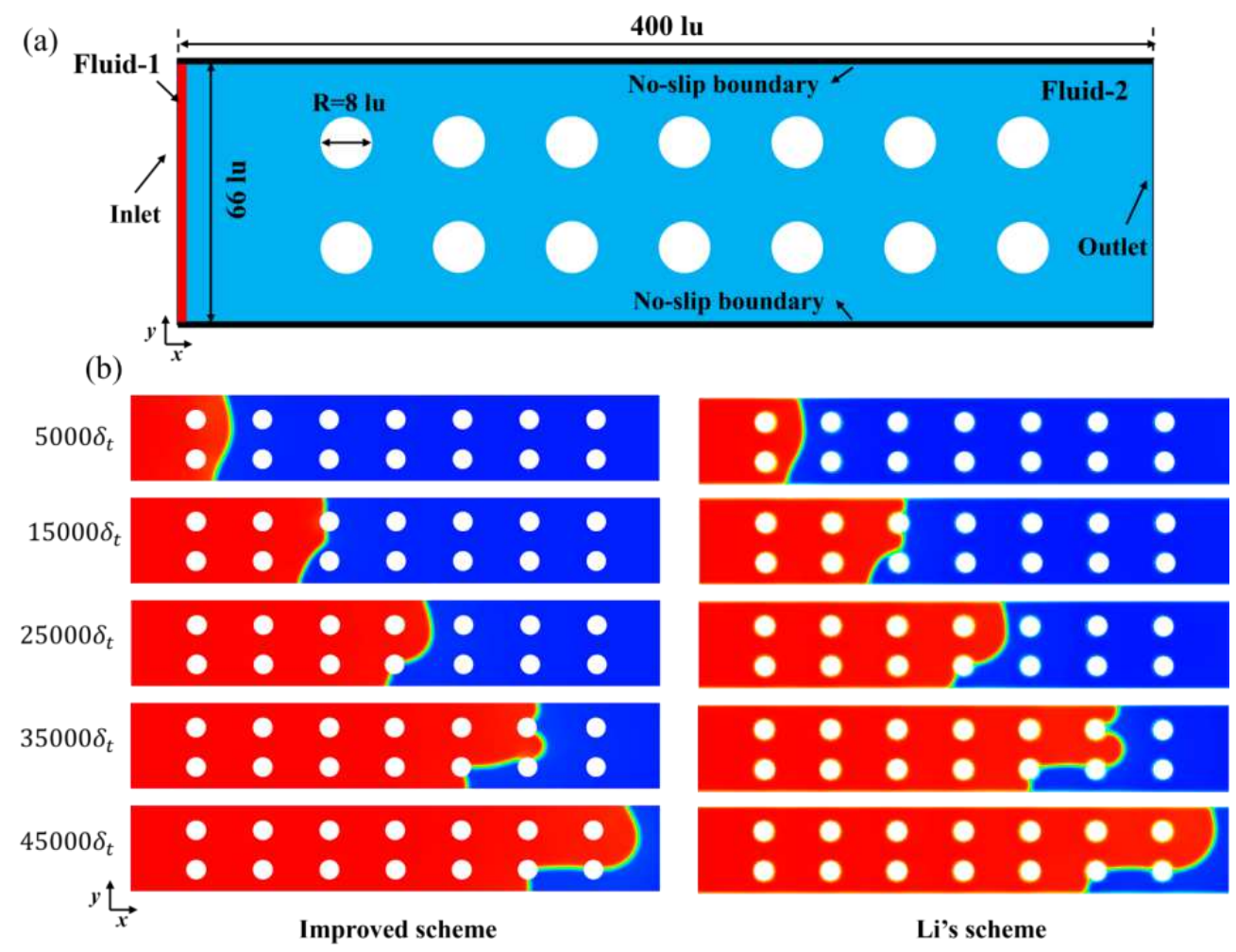

**Fig. 13.** Validation of intense non-uniform flow in porous media: (a) Schematic of the displacement process. (b) Displacement patterns simulated by improved solid-fluid scheme and Li's scheme[46].

## E. Contact Angles on Spherical Surface

To verify the generality of the improved solid-fluid scheme proposed in this paper for 3-D situations, we also have simulated 3-D static-contact angles on a spherical surface. The computational domain has 200×200×280 grids, with a solid cylinder of radius $R$=50 positioned at (100, 100, 100). A circular droplet (fluid 1, radius $r$ = 45) was placed at (100, 100, 180), and the remaining area was filled with fluid 2. The periodic boundary condition is applied in all directions, and the halfway bounce-back scheme is employed to treat the spherical surface. Physical parameters of two-phase fluid are the same as those used in the above two-dimensional tests. It should be noted that the improved scheme proposed in this paper can be applied to 3-D simulation without any modifications to the algorithm and implementation process. In terms of code implementation, only the dimension in the z-direction needs to be added. In addition, the 3-D model used in this simulation is non-orthogonal D3Q19, and the corresponding discrete velocity and weight coefficients are come from Li et al.[51]. Fig.14 presents the results of different three-dimensional contact angles obtained by the improved solid-fluid scheme and it agrees with the simulation results of Li's scheme[46]. This indicates that the improved solid-fluid scheme proposed in this paper can adapt well to 3-D scenes.

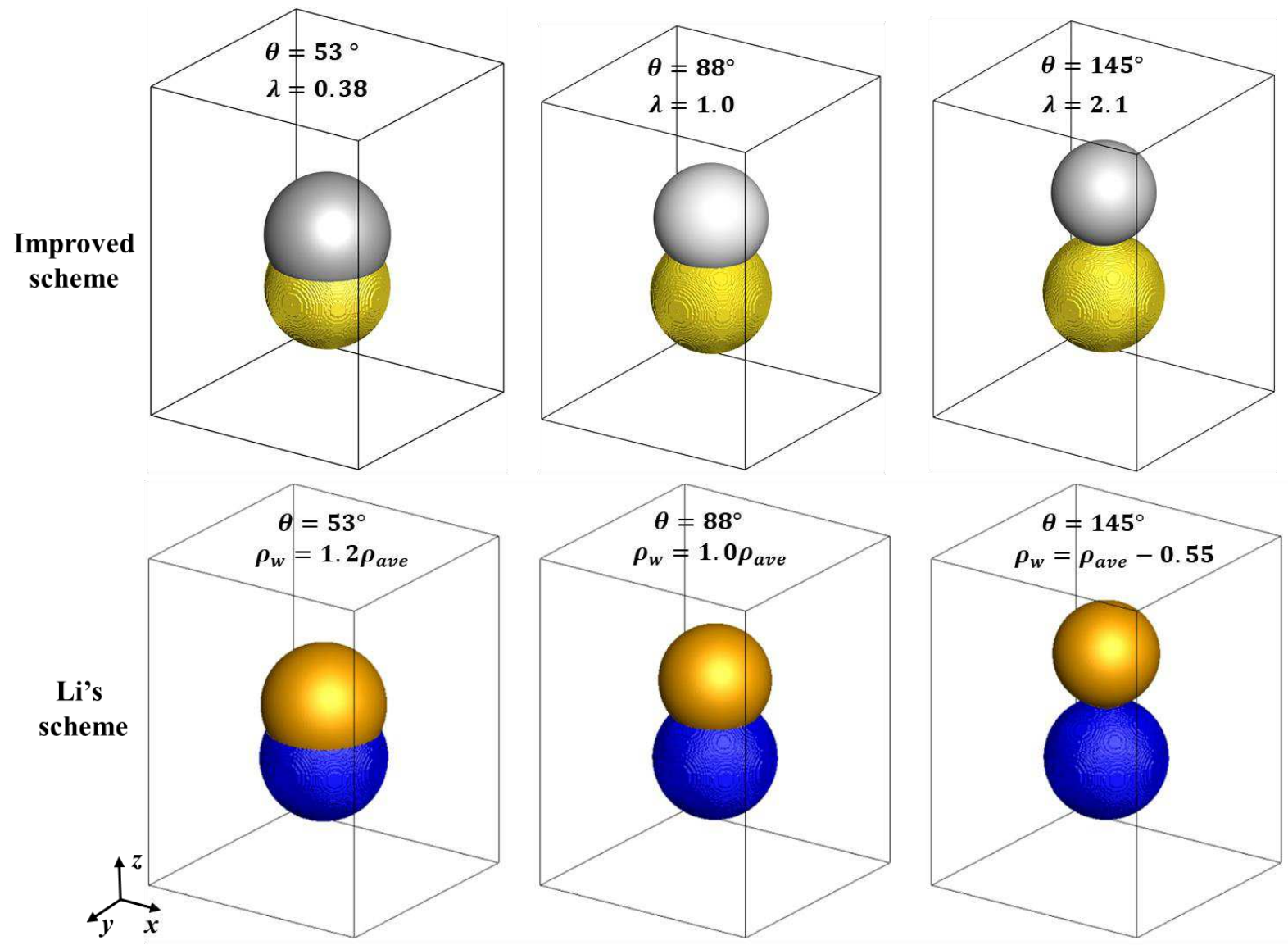

**Fig. 14.** Validation of 3-D static contact angle simulations on a sphere surface.

Finally, we summarize the characteristics of three contact angle schemes and list them in Table III. The comparative analysis reveals that while all three schemes have limitations, the improved scheme uniquely achieves the complete elimination of unphysical mass transfer layers across its wide operating range (10–165°), effectively addressing the core deficiency of the original scheme. Unlike Li's scheme[46], which still exhibits residual mass transfer artifacts at smaller contact angles, the improved scheme maintains physical fidelity with only localized spurious currents at contact points.

**Table III**. Comparison of three solid-fluid interaction schemes.

| Contact angle scheme | Equation | Contact angle range | Advantage | Shortcoming |
|---|---|---|---|---|
| Original[18] | $\boldsymbol{F}_{k_ads} = -G_{w,k}\rho_k(x)\sum_i w_i s(x+e_i\Delta t)\boldsymbol{e}_i$<br>$G_{w,k} = constant$ | 33-120° | Simple application | Spurious currents with obvious unphysical mass transfer layer |
| Improved | $\boldsymbol{F}_{k_ads} = -G_{w,k}\rho_k(x)\sum_i w_i s(x+e_i\Delta t)\boldsymbol{e}_i$<br>$G_{w,1} = \begin{cases} G_{w,1}, & \rho^e > \delta \\ g_1\ (\rho^e), & \delta \geq \rho^e > 0 \\ g_2\ (\rho^e), & 0 \geq \rho^e > -\delta \\ G_{w,2}, & \rho^e < -\delta \end{cases}$ | 10-165° | Effectively eliminate unphysical mass transfer layers | Large local spurious currents at contact point |
| Li's[46] | $\boldsymbol{F}_{k_ads} = -\rho_k(x)\sum_i w_i \rho_k^{'}(x+e_i\Delta t)\boldsymbol{e}_i$<br>$\rho_k^{'} = \begin{cases} \rho_w \cdot s(x+e_i\Delta t) \\ \rho_k^{'} \cdot (1-s(x+e_i\Delta t)) \end{cases}$<br>$\rho_w = \begin{cases} \Psi \cdot \rho_{ave}, & \theta \leq 90^\circ \\ (\rho_{ave} - \Delta\rho), & \theta > 90^\circ \end{cases}$<br>$\rho_{ave} = \frac{\sum_i w_i \rho_k(x+e_i\Delta t)s(x+e_i\Delta t)}{\sum_i w_i s(x+e_i\Delta t)}$ | 10-165° | Effectively reduce spurious currents | Unphysical mass transfer layer still exists at smaller contact angles |

## IV. Conclusion

The original solid-fluid interaction scheme, widely employed in pseudopotential models for simulating wetting phenomena, offers conceptual simplicity but suffers from critical limitations, including restricted contact angle ranges, spurious currents artifacts, and unphysical mass-transfer layers near boundaries. To address these challenges, this study proposes an improved solid-fluid interaction scheme that replaces constant adhesion parameters with localized interaction parameters governed by effective fluid densities. This modification significantly enhances mechanical stability near solid surfaces while preserving the scheme's computational simplicity. Key advancements include:

(1) Extended Applicability: The improved scheme achieves a broader spectrum of contact angles ($\theta \approx 10°–165°$) without unphysical mass-transfer layers, overcoming the original scheme's limitations ($\theta \approx 33°–120°$).

(2) Suppressed Spurious Currents: By dynamically adjusting adhesion forces based

on local fluid densities, spurious currents are reduced by 40–50% in hydrophilic regimes (θ < 90°) and 30–40% in hydrophobic regimes (θ > 90°).

(3) Robust Boundary Handling: An eighth-order isotropic interaction scheme coupled with a novel second-layer lattice reconstruction method ensures accurate simulations even at complex curved boundaries.

Validation through four benchmark cases—static contact angle, droplet dynamics through confined geometries, immiscible displacement, and co-current flow—demonstrates the versatility and accuracy of our scheme. Quantitative comparisons with analytical solutions and prior studies reveal relative errors below 5%, affirming its reliability for diverse multiphase flow scenarios. This work establishes a foundation for high-fidelity simulations of wettability-driven phenomena in pseudopotential LBM frameworks, with potential applications in energy systems, biomedical engineering, and environmental fluid dynamics.

## Acknowledgments

The authors acknowledge the financial support of the National Natural Science Foundation of China (No. U20A20299), the Natural Science Foundation of Guangdong Province (No: 2024A1515012393), and the Guangzhou Science and Technology Plan Project (No: 2025A04J7055).

## Appendix A: Algorithm

| Algorithm of Improved solid-fluid scheme |
| --- |
| // Calculate local solid-fluid interaction parameters ( $G_{w,1}(x, y)$) at each fluid node |
| 1: for each node $(x, y)$ do |
| 2: if (the node located at $(x, y)$ is a fluid node) then |
| 2: calculating $\rho^e(x, y)$ using Eq. (10) |
| 3: if ( $\rho^e(x, y) > \delta$ ) then $G_{w,1}(x, y) = G_{w,1}$ |

4: else if ( $\delta \geqslant \rho^e(x, y) > 0$) then $G_{w,1}(x, y) = g_1(\rho^e)$

5: else if ($0 \geqslant \rho^e(x, y) > -\delta$) then $G_{w,1}(x, y) = g_2(\rho^e)$

6: else if ( $\rho^e(x, y) < -\delta$) then $G_{w,1}(x, y) = G_{w,2}$

7: end if

8: end if

9: end for

// Calculate local solid-fluid interaction force ( $F_{k_ads}(x, y)$) at each fluid node

10: for each node $(x, y)$ do

11: if (the node located at $(x, y)$ is a fluid node) then

12: calculating $F_{k_ads}(x, y)$ using Eq. (9)

13: end if

14: end for

Source code link: shixuemozun/Improved-solid-fluid-scheme-LBM-SC-

## Appendix B: Detailed description of boundary scheme

For the treatment of non-slip conditions, this paper adopts half-bounce back scheme[52], represented by Eq. (14), which can effectively ensure the conservation of mass in the simulation system; For periodic boundaries[53], as shown in Eq. (15), this scheme is mainly used to handle infinitely long channels and can also ensure strict conservation of mass. For the velocity boundary of inlet and outlet, this paper adopts the Zou-He scheme[5], as shown in Eq. (16), which has high accuracy and good numerical stability in two-dimensional simulations.

$$\boldsymbol{f}_{i_}^{k}(\boldsymbol{r}_b, t+\Delta t) = \boldsymbol{f}_{i}^{*,k}(\boldsymbol{r}_b, t)\,. \tag{14}$$

$\boldsymbol{r}_b$ is the fluid boundary node, $\boldsymbol{f}_i^{*,k}$ is distribution function after collision, $i_$ and $i$ represent opposite discrete velocity directions, respectively.

$$\begin{aligned}
&\boldsymbol{f}_{1,5,8}^{*,k}(\boldsymbol{r}_{b_r}, t) = \boldsymbol{f}_{1,5,8}^{k}(\boldsymbol{r}_{b_l}, t+\Delta t)\\
&\boldsymbol{f}_{3,6,7}^{*,k}(\boldsymbol{r}_{b_l}, t) = \boldsymbol{f}_{3,6,7}^{k}(\boldsymbol{r}_{b_r}, t+\Delta t)\\
&\boldsymbol{f}_{2,5,6}^{*,k}(\boldsymbol{r}_{b_u}, t) = \boldsymbol{f}_{2,5,6}^{k}(\boldsymbol{r}_{b_d}, t+\Delta t)\\
&\boldsymbol{f}_{4,7,8}^{*,k}(\boldsymbol{r}_{b_d}, t) = \boldsymbol{f}_{4,7,8}^{k}(\boldsymbol{r}_{b_u}, t+\Delta t)
\end{aligned}. \tag{15}$$

$\boldsymbol{r}_{b_u}$, $\boldsymbol{r}_{b_d}$, $\boldsymbol{r}_{b_l}$ and $\boldsymbol{r}_{b_r}$ respectively represent the upper, down, left, and right boundaries.

$$\begin{cases}
\rho^k = \sum_{i=0}^{8} f_i^k\\
f_1^k + f_5^k + f_8^k - f_3^k - f_6^k - f_7^k + 0.5\cdot(F_{k_in,x} + F_{k_ads,x}) = \rho^k \cdot u_x^{eq}\\
f_2^k + f_5^k + f_6^k - f_4^k - f_7^k - f_8^k + 0.5\cdot(F_{k_in,y} + F_{k_ads,y}) = \rho^k \cdot u_y^{eq}
\end{cases}. \tag{16}$$

$F_{k_in,x}$ and $F_{k_in,y}$ represent the fluid-fluid interaction forces in the $x$ and $y$ directions, respectively. $F_{k_ads,x}$ and $F_{k_ads,y}$ represent the solid-fluid interaction forces in the $x$ and $y$ directions, respectively. For the inlet boundary, after streaming, the distribution functions $f_1^k$, $f_5^k$, and $f_8^k$ are the variables to be solved, while the rest are known variables; For the outlet boundary, after streaming, the distribution functions $f_3^k$, $f_6^k$, and $f_7^k$ are the variables to be solved, while the rest are known variables.

## Appendix C: Dimensional conversion

Table IV. Dimensional conversion (example for the case in Sec III B).

| Physical property | SI units | Conversion factor | Lattice units |
|---|---|---|---|
| Length | $5\times10^{-4}$ m | $C_l = \frac{5\times10^{-4}}{500} = 10^{-6}$ | 500 lu |
| Density | $10^3$ kg/m$^3$ | $C_m = \frac{10^3}{1.0}\cdot C_l^3 = 10^{-15}$ | $\rho_1 = \rho_2 = 1.0$ mu/lu$^3$ |
| Time step | $1.67\times10^{-7}$ s | $C_t = \frac{C_l^2}{C_\upsilon} = 1.67\times10^{-7}$ | 1 ts |
| Kinematic viscosity | $10^{-6}$ m²/s | $C_\upsilon = \frac{10^{-6}}{0.167} = 5.98\times10^{-6}$ | $\upsilon_1 = \upsilon_2 = 0.167$ lu²/ts |
| Surface tension | $13.5\times10^{-3}$ kg/s² | $C_\gamma = \frac{C_m}{C_t^2} = 0.036$ | $\gamma = 0.375$ mu/ts² |
| Velocity | $2.39\times10^{-3}$ m/s | $C_v = \frac{C_l}{C_t} = 5.98$ | $4\times10^{-4}$ lu/ts |

## Appendix D: Experimental details

In this contact angle measurement experiment, the measuring instrument used is OCA15Pro, as shown in Fig.15. Electric syringe unit ES integrated within the contact angle measuring instrument was utilized to precisely dispense liquid droplets onto polydimethylsiloxane (PDMS) substrates that had been surface-modified with either hydrophilic (XN-534) or hydrophobic (XN-532) reagents, thereby generating varying contact angle profiles. Subsequently, the real-time images captured by the CCD camera were analyzed using the SCA20 image processing software to measure the contact angle.

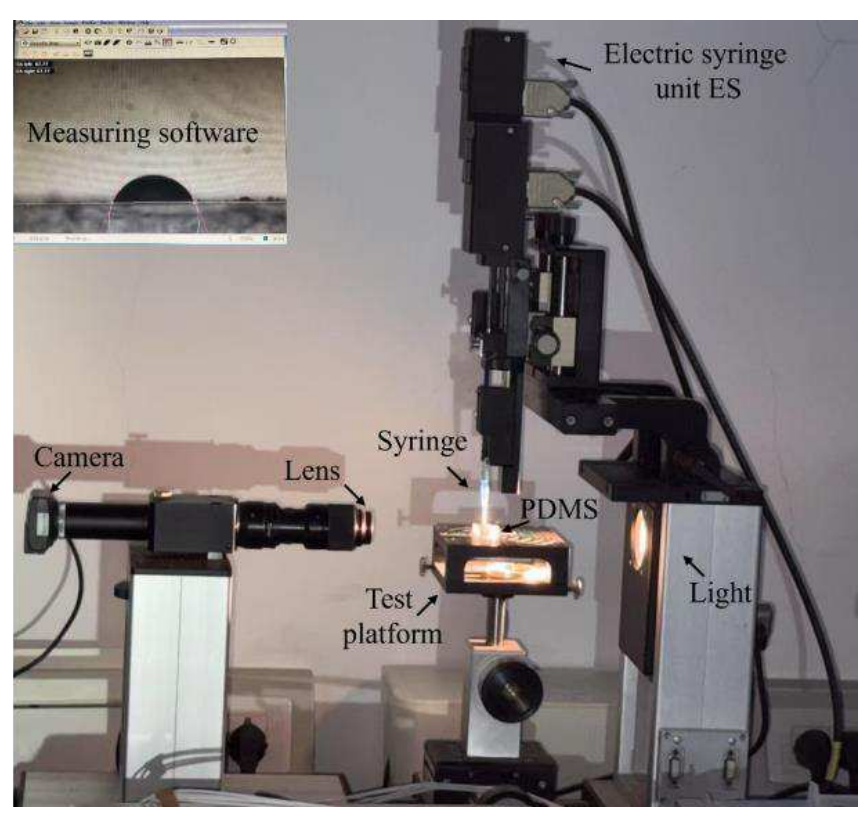

**Fig. 15.** Contact angle measuring instrument: OCA15Pro.

Table V. Detailed information about observation equipment (OCA15Pro).

| System module | Parameter details | Resolution/accuracy |
|---|---|---|
| Imaging system | Sensor: 1/2 "CCD, black and white<br>Lens: 100mm focal length, 6x optical | 768 × 576 pixels |
| | magnification | |
| | Light source: halogen lamp | |
| | Software: SCA20 | |
| Measuring software | Measurement range: 0~180 ° | ± 0.1 ° |
| | Fitting algorithm: sessile drop method | |
| | Control mode: Electric syringe unit ES | |
| Droplet system | Syringe specification: 1ml | |
| | Needle inner diameter: 0.2mm | |